\documentclass[aps,prd,reprint,groupedaddress]{revtex4-2}
\input epsf.sty
\usepackage{graphicx,amsmath,amssymb}
\def\lromn#1{\uppercase\expandafter{\romannumeral#1}}

\begin{document}

% Use the \preprint command to place your local institutional report
% number in the upper righthand corner of the title page in preprint mode.
% Multiple \preprint commands are allowed.
% Use the 'preprintnumbers' class option to override journal defaults
% to display numbers if necessary
%\preprint{magnetization-renp}

%Title of paper
\title{
New types of instability and CP violation in electroweak theory
}

% repeat the \author .. \affiliation  etc. as needed
% \email, \thanks, \homepage, \altaffiliation all apply to the current
% author. Explanatory text should go in the []'s, actual e-mail
% address or url should go in the {}'s for \email and \homepage.
% Please use the appropriate macro foreach each type of information

% \affiliation command applies to all authors since the last
% \affiliation command. The \affiliation command should follow the
% other information
% \affiliation can be followed by \email, \homepage, \thanks as well.

\author{M. Yoshimura}
%\affiliation{Okayama University}
\affiliation{Research Institute for Interdisciplinary Science,
Okayama University \\
Tsushima-naka 3-1-1 Kita-ku Okayama
700-8530 Japan}

%Collaboration name if desired (requires use of superscriptaddress
%option in \documentclass). \noaffiliation is required (may also be
%used with the \author command).
%\collaboration can be followed by \email, \homepage, \thanks as well.
%\collaboration{}
%\noaffiliation

\date{\today}

\begin{abstract}
% insert abstract here
It is known that the Schwinger mechanism in vector-like QED theory 
is afflicted by a logarithmic singularity for a hypothetical massless charged  fermion.
We extend singularity analysis to a more realistic case of the chiral electroweak theory,
to show that the effective lagrangian under background gauge field at zero temperature 
exhibits a similar instability proportional to $\ln (1/m_{\nu}^2)$ with
$m_{\nu}$ a small neutrino mass.
Moreover, the effective lagrangian of chiral fermion loop
contains CP violating pieces  proportional to background gauge fields
in odd powers  of $\vec{E}_Z\cdot\vec{B}_Z$ or 
$(\vec{E}_{W^+}\cdot\vec{B}_{W^-}+ \vec{E}_{W^-}\cdot\vec{B}_{W^+})/2 $. 
This brings in a  new source of CP violation and time-reversal symmetry violation
in the standard particle theory
independent of the Kobayashi-Maskawa phase of quark mass mixing matrix.
The effective action in thermal equilibrium at finite temperature $T$ is then calculated
under background SU(2)$\times $U(1) gauge fields in the 
spontaneously broken phase.
An even more singular power-law behavior $\propto (m_{\nu} T)^{-5/2}$ 
 is found and it contains CP violating term as well.
The case of Majorana neutrino satisfies almost all necessary conditions
to generate a large lepton number asymmetry, though not
necessarily convertible to a baryon asymmetry due to lower cosmic
temperatures at which this may occur.

\end{abstract}

% insert suggested keywords - APS authors don't need to do this
%\keywords{}

%\maketitle must follow title, authors, abstract, and keywords
\maketitle

% body of paper here - Use proper section commands
% References should be done using the \cite, \ref, and \label commands

% Put \label in argument of \section for cross-referencing
%\section{\label{}}
%\subsection{}
%\subsubsection{}

\section{Introduction} 
Instabilities of vector-like gauge theories are well known in 
QCD (Quantum ChromoDynamics) and  in QED (Quantum ElectroDynamics).
QCD vacuum instability is understood by instanton effects  in
Euclidean field theory \cite{coleman lecture}.
This led to discovery of the  strong CP violation problem,
giving its possible resolution by  a chiral symmetry
known as Peccei-Quinn symmetry \cite{pq}.
Instability in QED is known as 
Schwinger mechanism \cite{schwinger} 
related to $e^{\pm}$ pair production.
Non-abelian extension of Schwinger mechanism
\cite{yildiz-cox}, \cite{savvidy}, \cite{claudson-yildiz-cox} 
has deepened our understanding of instability:
non-abelian instability is related to the magnetic property of
vector gauge boson \cite{no} believed to be the origin of
asymptotic freedom as well.

Effects we discuss in the present work are different:
it works in chiral gauge theories such as the electroweak theory.
The proper-time background method used in \cite{schwinger}, \cite{yildiz-cox}, 
\cite{claudson-yildiz-cox} is powerful
enough to be extended to the electroweak  gauge field theory,
as shown in the present work.
Results in chiral theory can be derived from
a simple transformation rule from the corresponding vector-like gauge theory.

Results of physical relevance we derive are a surprise:
we find a new source of CP violation and time-reversal symmetry violation.
This adds to the well-known Kobayashi-Maskawa phase of CP violation,
and is more dynamical.
The present work is limited to our fundamental findings and many possible applications
are relegated to future works.

The rest of this paper is organized as follows.

=============================

%\section
\lromn2 \; %\hspace{0.3cm}
{\bf The proper-time background field method: a summary} 

\vspace{0.3cm}
\lromn3 \; %\hspace{0.3cm}
{\bf  Chiral SU(2) $\times$ U(1) electroweak theory}

{\bf A. Chiral theory from vector-like theory}

{\bf B. Flavor dependence of gauge coupling}

{\bf C. Maximum CP violation under gauge field background}

{\bf D. Light neutrino one-loop contribution}

\vspace{0.3cm}
\lromn4 \; %\hspace{0.3cm}
{\bf Electroweak theory under field background at finite temperature} 

{\bf A. Effective action under background gauge fields at finite temperature}

{\bf B. Massless limit at finite temperature}

{\bf C. High temperature limit and a mechanism of generating lepton number asymmetry}

{\bf D. Speculation on a possible phase transition}

\vspace{0.3cm}
\lromn5 \; %\hspace{0.3cm}
{\bf Summary and outlook}

\vspace{0.3cm}
\lromn6 \; %\hspace{0.3cm}
{\bf Appendix: Additional notes} 

{\bf A.
Logarithmic singularity in the massless fermion limit}

{\bf B. Color matrix eigenvalues}

{\bf C.
Scalar and gauge boson loops}

{\bf D.
General four-component neutral lepton system}

=============================

Throughout this work we shall use the natural unit of
$\hbar = c =1$, and the Boltzmann constant $k_B = 1$
unless otherwise stated.

\section{The proper-time background field method: a summary}

We recapitulate the proper time formalism of QED under
constant electric and magnetic background field denoted by
$\vec{E}\,, \vec{B}$.
Heisenberg-Euler non-linear QED \cite{heisenberg-euler}  along with
$e^{\pm}$ pair production rate is derived in this formalism.

The starting point in QED is
the electron one-loop effective lagrangian (more properly
lagrangian density) prior to renormalization,
\begin{eqnarray}
&&
{\cal L}^{(1)} (x) = \frac{i}{2} \int_0^{\infty} \frac{ds}{s}\, e^{-  m_e^2 s}
{\rm tr} \langle x | U(s) | x \rangle
\,,
\label {unrenormalized lagrangian}
\\ &&
U(s) = e^{- i {\cal H} s}
\,, \hspace{0.5cm}
{\cal H} = (p_{\mu} -  A_{\mu})^2 - \frac{1}{2} \sigma_{\mu\nu} F^{\mu \nu}
\,.
\end{eqnarray}
${\cal H}$ is the quadratic form of Dirac hamiltonian in the proper-time
formalism.
We include the gauge coupling constant $e$ in the definition of
$A_{\mu}\,, \vec{E}\,, \vec{B}$ to make non-abelian generalization easier.
This formula encompasses all even orders of background fields $F_{\mu\nu}$
in external lines, odd powers being forbidden by the Furry's theorem.
One can extract contribution from a specific powers of fields
by a functional derivative of this formula.
The attractive feature  is that the formula is written in terms
of gauge invariant quantity, since one may use c-number $k_{\mu} = p_{\mu} - A_{\mu}$.

Since there is no field operator involved in the above formula, 
c-number eigenvalues of matrix $ \sigma_{\mu\nu} F_{\mu \nu}$ can be used.
Noting the identity, 
\begin{eqnarray}
&&
( \frac{1}{2} \sigma_{\mu\nu} F^{\mu\nu})^2 
= 2 (- {\cal F} + i \gamma_5\, {\cal G} )
\,, 
\label {fmn identity}
\\ &&
{\cal F} = \frac{1}{2} (\vec{E}^2 - \vec{B}^2)
\,, \hspace{0.5cm}
 {\cal G} = \vec{E}\cdot \vec{B}
\,,
\label {f_ab eigenvalues}
\end{eqnarray}
one derives four eigenvalues of $\frac{1}{2} \sigma_{\mu\nu} F_{\mu \nu} $,
\begin{eqnarray}
&&
\pm \left( 2(- {\cal F} \pm i {\cal G} )\right)^{1/2}
\,.
\end{eqnarray}
Schwinger uses one of these, 
\begin{eqnarray}
&&
X =  \left( 2(- {\cal F} + i {\cal G} )\right)^{1/2}
\,,
\end{eqnarray}
others being related by complex conjugation and
sign reversal.

Extension of Schwinger's U(1) gauge theory to non-abelian gauge theories
has been completed by many.
We mention works of \cite{yildiz-cox} $\sim$  \cite{claudson-yildiz-cox}
for convenience of our subsequent discussion.
In non-abelian extensions U(1) gauge invariants are generalized to
\begin{eqnarray}
&&
{\cal F} = \frac{1}{2} \sum_a t_a (\vec{E}_a^2 - \vec{B}_a^2)
\,, \hspace{0.3cm}
{\cal G} = \sum_a t_a \vec{E}_a\cdot\vec{B}_a
\,,
\end{eqnarray}
with the index $a$ running over independent gauge degrees of freedom.
$t_a$'s are coefficients giving eigenvalues of non-abelian adjoint matrix diagonalization,
as described in Appendix B.
In electroweak SU(2) case they are $\pm 1/2$, and one can effectively 
take out $t_a$ factors in the above definition.
For SU(3) the problem is more complicated. See Appendix B for this.

A problem of the Schwinger variable $X$ is coexistence of CP even and odd terms:
${\cal F}$ is CP-even and ${\cal G}$ is CP-odd.
This makes application of this formalism to chiral theories
difficult, in particular in clarifying CP properties of its effective action.
We shall use other variables used in \cite{yildiz-cox}, 
 \cite{claudson-yildiz-cox}.
Including coupling factors and introducing non-abelian indexes,
these are two variables $a\,, b$ defined by
\begin{eqnarray}
&&
a^2({\cal F}, {\cal G}) =  {\cal F} + \sqrt{{\cal F}^2
+  {\cal G}^2} 
\,,
\\ &&
b^2({\cal F}, {\cal G}) = - {\cal F} + \sqrt{{\cal F}^2
+  {\cal G}^2} 
\,.
\end{eqnarray}
Both of these, $a^2, b^2$, are CP-even, but their
square-root product $a b ={\cal G} $
is CP-odd, when the square-root cut singularity is analytically continued.

The better formula of one-loop fermion contribution, after renormalization, is given by
\begin{eqnarray}
&&
\Delta {\cal L}_f = - \frac{1}{8 \pi^2} \int_{0 + i 0^+}^{\infty + i 0^+}
\frac{ds }{s^3} e^{- m^2 s} {\cal M}_f (a, b; s)
\,,
\label {qed in a,b}
\\ &&
{\cal M}_f (a, b; s) =
 s^2 \, a b \cot( sa) \coth (s b) - 1 + \frac{s^2}{3} (a^2 - b^2)
\,.
\nonumber \\ &&
\label {qed mf}
\end{eqnarray}
When this formula is restricted to the abelian case,
it gives equivalent lagrangian to QED's.
This formula contains both real and imaginary parts.
The imaginary part may be derived by taking
$ \delta {\cal L}_f - (\delta {\cal L}_f )^* $.
This difference gives an infinite discrete series of contour integrations
around poles (arising from $\cot (sa)$)
on the imaginary axis, at $ s = i n \pi/a\,, n=1,2, \cdots$.
The total imaginary part is \cite{claudson-yildiz-cox}
\begin{eqnarray}
&&
\Im \Delta {\cal L}_f = \frac{1}{8 \pi^2} ab 
 \sum_{n=1}^{\infty} \frac{1}{n} e^{- m^2 n \pi/|a| } 
\coth \frac{b n \pi }{ a} 
\,.
\label {imaginary part of f-loop}
\end{eqnarray}

Despite of the explicit appearance $a b= {\cal G}$ as an overall factor
$ a b \cot( sa) \coth (s b)$
and apparent evenness in ${\cal G}$ of $a$ and $b$,
the function ${\cal M}_f (a, b; s) $ is  even in the variable 
$a b = {\cal G}$,
viewed as two-variable function, $ab$ and $ a^2 - b^2= {\cal F}$.
This ensures that the effective action is CP conserving.
On the other hand, the formula (\ref{qed in a,b}) contains both even and odd power terms
of $a^2-b^2$, which becomes important later in 
application to chiral theories.
This property of ${\cal G}$ even function
 is due to that the analytic structure of this function around
the double zero ${\cal G} =0\,, {\cal F} =0$:
there are branch cut singularities of the square root type
along the imaginary axis at $\Im {\cal G} = 0 \sim - i \infty\,,
\Im {\cal F} = 0 \sim - i \infty$.
We have verified  by using the power series expansion
of a computer software
the property that ${\cal F}-$odd terms appear in total odd powers
of ${\cal F}$ and ${\cal G}$.
Total even power terms contain summed products of even ${\cal F}$ and even
${\cal G}$ powers.

Expansion of ${\cal M}_f (a, b; s) $ in power series in $s$ 
is equivalent to field power
expansion.
The first few expansion terms are
\begin{eqnarray}
&&
{\cal M}_f (a, b; s) = - \frac{1}{45} (4  {\cal F}^2 + 7{\cal G}^2  ) s^4
\nonumber \\ &&
- \frac{2}{945}  {\cal F} (8  {\cal F}^2 + 13 {\cal G}^2  ) s^6 + O(s^8)
\,.
\label {two leading contributions to s-integrand}
\end{eqnarray}
The expansion contains even $s$ powers,
 and  $s^{ 2 (2n +1)}$ terms contain odd ${\cal F}$ powers,
and all other power terms are even in both variables.
$s-$integration over the first term gives the well-known
Heisenberg-Euler non-linear QED lagrangian \cite{heisenberg-euler} .

The infrared limit of massless fermion becomes of great interest
in chiral theories due to small neutrino masses, and it would be instructive to work out
the massless electron limit although this is a fictitious setting.
It is easy to explain the logarithmic behavior of the imaginary part given by
(\ref{imaginary part of f-loop}).
Assuming that the interchange of the limit and the sum in this equation is allowed,
the large $n$ behavior of the sum is of order $1/n$,
which gives a divergence.
On the other hand, a finite mass $m$ gives a well-defined convergent result,
implying a logarithmic singular behavior.
Indeed, according to calculation in Appendix A, the massless limit gives
\begin{eqnarray}
&&
\Im \Delta {\cal L}_f \rightarrow \frac{ab}{8 \pi^2} \ln \frac{ |a| t_0}{m^2 \pi }
\,,
\end{eqnarray}
with a factor $t_0$ of order unity not precisely determined.
The real part of effective lagrangian exhibits the logarithmic
singularity as well.

The color or flavor
matrix diagonalization in the standard theory of particle physics
is given in Appendix B.
Non-abelian gauge boson and scalar one-loop contributions
have been worked out in the literature,
and we list  in Appendix C functions,
${\cal M}_g\,, {\cal M}_s$ that correspond to the fermion ${\cal M}_f$.

\section{Chiral SU(2) $\times$ U(1) electroweak theory}

We discuss in the present work the gauge invariant effective action under
non-abelian gauge field background, which automatically selects
an even number of external gauge fields.
This circumstance makes the following trick to work.

\subsection{Chiral theory from vector-like theory}

Chiral gauge theories are theories in which fermions have gauge coupling
asymmetric in the left-handed current proportional to $\gamma (1-\gamma_5)/2$
and the right-handed current $\propto \gamma (1+\gamma_5)/2$.
The  most important case
is the electroweak theory in which the SU(2) triplet gauge boson $\vec{W}$
couples to the left-handed chiral fermion and the singlet gauge boson $B$
couples to  fermions of both chirality,
with different gauge coupling constants, $g\,, g'$ via covariant 
derivatives,
\begin{eqnarray}
&&
D_{\mu} = \partial_{\mu} - i \frac{g}{2} \vec{W}_{\mu} \cdot \vec{\tau} 
- i g' Y B_{\mu}
\,,
\end{eqnarray}
with the hypercharge $Y$ assigned to quarks and leptons, separately.
These are
$Y = Y_L^q = 1/6$ for quark doublets, $Y_L^l = -1/2$ for lepton doublets,
and all different numbers for  right-handed $u_R, d_R, \nu_R, e_R$ SU(2) singlets.
For the moment we assume  neutrinos of finite Dirac type masses.
We work out results in the broken phase, thus all fermions and
gauge bosons except the photon are massive.

Charge neutral gauge field ($Z$ and $\gamma$) couplings are thus summarized as an ordinary 
electromagnetic coupling to photon (with $e\, Q_f $ charges of fermion $f$)
and Z-boson coupling of the well-known form,
\begin{eqnarray}
&&
Z \cdot \gamma \, \sqrt{g^2+ (g')^2} \left( T_3 L - \sin^2 \theta_w Q_f (L + R)
\right)  
\,,
\end{eqnarray}
with $T_3 = \sigma_3/2$ SU(2) quantum numbers and $(L, R)= (1 \mp \gamma_5)/2$.
It is important to keep in mind that each chiral component 
maintains the same chiral property in all orders of couplings.
For convenience we call the contribution $\propto \gamma_5$ the axial-vector contribution,
and the one without it the vector contribution.

We now derive  a general replacement rule that the one-loop contribution
proportional to $\gamma_5$ 
is derived from the contribution without $\gamma_5$.
Consider numerator factor in perturbation expansion of $2n$ orders
under gauge field background.
In perturbation theory background fields are given by vector potentials,
and we give input momenta to vector fields, later taken to recover the limiting
constant field strength  of zero momentum.
Decomposition into chiral components shows that mass terms drop out,
and the probability amplitude is 
\begin{eqnarray}
&&
{\rm tr} A_1 \cdot \gamma \frac{1 \mp \gamma_5}{2} k_1 \gamma \,
A_2 \cdot \gamma \frac{1 \mp \gamma_5}{2} k_2 \gamma
\cdots
A_{2n} \cdot \gamma \frac{1 \mp \gamma_5}{2} k_{2n} \gamma 
\nonumber \\ &&
=
{\rm tr} A_1 \cdot \gamma  k_1\,  \gamma  A_2 \cdot \gamma  k_2 \gamma \cdots
A_{2n} \cdot \gamma  k_{2n} \gamma \frac{1 \pm \gamma_5}{2}
\,,
\end{eqnarray}
where $A_i$'s are external gauge fields, and $k_i$ are input momenta.
This shows the amplitude relation in chiral theories 
such that the axial-vector contribution is related to
the vector contribution,
\begin{eqnarray}
&&
{\rm tr} \langle x | {\cal H}^{2n}_A | x \rangle
= \pm {\rm tr} \langle x | {\cal H}^{2n}_V \gamma_5 | x \rangle
\,,
\end{eqnarray}
in expansion of (\ref{unrenormalized lagrangian}).

In the proper-time formalism 
the eigenvalue formula (\ref{fmn identity}) may be used to relate
the axial-vector contribution to the vector-relation.
To extract the axial-vector contribution, note
\begin{eqnarray}
&&
\left( (p_{\mu} - A_{\mu} )^2 - \frac{1}{2} \sigma_{\mu \nu}F^{\mu \nu} \right)^{n}
\nonumber \\ &&
=
\left( (p_{\mu} - A_{\mu} )^2 \mp \sqrt{2} \sqrt{- {\cal F}_V + i \gamma_5 {\cal G}_V }\right)^{n}
\\ &&
\equiv {\cal H}^{2n}_V + \gamma_5 {\cal H}^{2n}_A
\,.
\end{eqnarray}
The axial-vector and the vector contributions contain numerator factors,
\begin{eqnarray}
&&
{\rm tr} \langle x | {\cal H}^{2n}_V|x \rangle =  - {\cal F}_V\,
{\rm tr} \langle x | {\cal H}^{2(n-1)}|x \rangle 
\,,
\\ &&
{\rm tr} \langle x | {\cal H}^{2n}_V|x \rangle = i  {\cal G}_V\,
{\rm tr} \langle x | {\cal H}^{2(n-1)}|x \rangle
\,.
\end{eqnarray}
Thus, the
vector and axial-vector contributions are  interchanged by
the rule, $-{\cal F}_V \leftrightarrow i {\cal G}_V$ in vector-like theories.

Thus, two integrand functions, vector and axial-vector
functions given using variables ${\cal F}, {\cal G}$,  are related by
\begin{eqnarray}
&&
{\cal M}_A^{L,R} ( {\cal F}\,, {\cal G}) = \mp {\cal M}_V^{L,R} ( - i {\cal G}\,, i {\cal F})
\,,
\label {axial from vector}
\\ &&
{\cal M}^L ( {\cal F}\,, {\cal G}) =  {\cal M}_V^{L} ( {\cal F}\,, {\cal G})
-  {\cal M}_V^{L} ( - i {\cal G}\,, i {\cal F})
\,,
\\ &&
{\cal M}^R ( {\cal F}\,, {\cal G}) =  {\cal M}_V^{R} ( {\cal F}\,, {\cal G})
+  {\cal M}_V^{R} ( - i {\cal G}\,, i {\cal F})
\,.
\end{eqnarray}
It becomes possible for the axial-vector contribution
to contain ${\cal G}$ odd terms, since the vector contribution
contains ${\cal F}$ odd terms.
In the example given by (\ref{two leading contributions to s-integrand}),
$s^6$ power terms have ${\cal F}$ odd powers, hence
the corresponding axial-vector contribution is
\begin{eqnarray}
&&
2 {\cal M}_A ( {\cal F}\,, {\cal G}) =
  \frac{1}{45} (4  {\cal G}^2 + 7{\cal F}^2  ) s^4
\nonumber \\ &&
- i \frac{2}{945}  {\cal G} (8  {\cal G}^2 + 13 {\cal F}^2  ) s^6 + O(s^8)
\,.
\label {s6 axial contribution}
\end{eqnarray}
The second term $\propto {\cal G}$ in this equation is CP violating (CPV).
When this is integrated over $s-$variable,
it gives CPV effective lagrangian of the form,
\begin{eqnarray}
&&
i \frac{ {\cal G} }{630  \pi^2} \,
  \, \frac{ (8  {\cal G}^2 + 13 {\cal F}^2  ) }{m^8}
\,.
\label {leading expanded cpv}
\end{eqnarray}
Formulas here are valid for small $({\cal F}, {\cal G})/m^4 $.
CPV terms are much larger at field regions of $({\cal F}, {\cal G}) = O(m^4) $.
Thus, it does not imply emergence of $1/m^8$ singularity.
We shall discuss the true massless limit in Subsection D.

\subsection{Flavor dependence of gauge coupling}

We shall first recover gauge coupling constants in field strength ${\cal F}\,, {\cal G}$,
which was deleted so far for simplicity.
These gauge constants differ depending on background gauge fields.
Under $W^{\pm}$ backgrounds alone (namely without mixed ones
such as $W^+W^- Z^2$), charged fermions contribute with the single
gauge coupling constant  $\pm g/2$, 
multiplied by  $T_3$ quantum numbers $\pm 1/2$.
Since even number of fields are relevant, all charged fermions
contribute with the same factor $(g/2)^2$.
Gauge coupling  to photon background alone is also simple:
$e Q_f$ with $Q_f$ the charge of each fermion $f$
in the unit of proton charge is the relevant one.

We next work out the effective action under Z-boson background
in the spontaneously broken phase, which is sensitive to details of flavor content.
There are six types of fermions with their chiralities specified circulating
loops:
$\nu_L\,, e_L\,, e_R\,, u_L\,, u_R\,, d_L\,, d_R$.
Their coupling factors are
\begin{eqnarray}
&&
\nu_L \, ; \frac{1}{2} = 0.5
\,, \hspace{0.3cm}
e_L\,; - \frac{1}{2} + 2 \sin^2 \theta_w \sim - 0.04
\,,
\label {z f-factor 1}
\\ &&
e_R\, ; \sin^2 \theta_w \sim 0.23
\,, \hspace{0.3cm}
u_L \,; \frac{1}{2} - \frac{2}{3} \sin^2 \theta_w \sim 0.35
\,,
\\ &&
u_R \,; - \frac{2}{3} \sin^2 \theta_w \sim - 0.15
\,, 
\\ &&
d_L\,; -\frac{1}{2} + \frac{1}{3} \sin^2\theta_w \sim - 0.42
\,, \hspace{0.3cm}
d_R\,; \frac{1}{3} \sin^2 \theta_w \sim 0.077
\,.
\nonumber \\ &&
\label {z f-factor 4}
\end{eqnarray}
where numerical values are obtained with $\sin^2 \theta_w = 0.23$
(close to the experimental value).
There is a wide variety of numerical range.
There are two more of the second and the third generations,
these contributions being characterized by their different masses.
We denote these factors by $g_{L, R}^f$ or $f_{L,R}= g_{L, R}^f/\sqrt{g^2 + (g')^2}$.

Flavor dependence of  CP conserving part is given by
\begin{eqnarray}
&&
\hspace*{-0.5cm}
\Re \Delta L_Z^{\rm CPC} = - \frac{1}{120 \pi^2} \frac{{\cal F}_Z^2 - {\cal G}_Z^2}{m^4}
\sum_f \left( (g_L^f)^4 + (g_R^f)^4
\right)
\,.
\end{eqnarray}
This lagrangian is manifestly real.

\subsection{Maximum CP violation under gauge field background}

It is  possible to separate CPV terms by a sort of projection, to derive
\begin{eqnarray}
&&
{\cal M}^{\rm CPV} ({\cal F}, {\cal G} ; t) \equiv \frac{1}{2} \left(
{\cal M}_A ({\cal F}, {\cal G} ; t ) - {\cal M}_A ({\cal F},- {\cal G} ; t)
\right)
\,,
\nonumber \\ &&
\label {cpv m-function 1}
\\ &&
{\cal M}^{\rm CPV} ({\cal F}, {\cal G}; t) = 
i t^2 {\cal F} \times
\nonumber \\ &&
\hspace*{-0.3cm}
\frac{\sin \frac{ t c}{\sqrt{2} } \cos \frac{ t d}{\sqrt{2} }  
\sinh \frac{ t c}{\sqrt{2} }  \cosh \frac{ t d}{\sqrt{2} }  
- ( c \leftrightarrow d)
 }
{( \cos \frac{ t c }{\sqrt{2} } \sinh \frac{ t c}{\sqrt{2} }
- i  \sin \frac{ t c }{\sqrt{2} } \cosh \frac{ t c}{\sqrt{2} }) ( c \rightarrow d) }
\nonumber \\ &&
+ \frac{ t^2}{2 } (c^2 - d^2)
\,,
\label {cpv m-function 2}
%\nonumber \\ &&
\\ &&
c^2({\cal F}, {\cal G} ) = - {\cal G} + \sqrt{{\cal F}^2 + {\cal G}^2}
\,, 
\label {cpv m-function 3}
\\ &&
d^2({\cal F}, {\cal G} ) =  {\cal G} + \sqrt{{\cal F}^2 + {\cal G}^2}
\,.
\label {cpv m-function 4}
\end{eqnarray}
In these formulas gauge coupling factors are included in
${\cal F}\,, {\cal G}$, and they can be readily recovered.
Renormalization subtraction is made so that the renormarlized lagrangian
contains no CPV term proportional to ${\cal G}$.
When expanded in powers of the integration variable $t$,
it gives the result,
\begin{eqnarray}
&&
{\cal M}^{\rm CPV} ({\cal F}, {\cal G}; t) =  -  \frac{i}{12} (c^4 - d^4) t^4 + O(t^6)
\,,
\\ &&
\Delta {\cal L}^{\rm CPV} ({\cal F}, {\cal G}; t) = - \frac{i}{24 \pi^2}
\frac{ {\cal G} \sqrt{{\cal F}^2 + {\cal G}^2}}{ m^4}
\,.
\end{eqnarray}

The contributions we calculated so far arises from the principal part integral
around $t=0$.
A more important contribution to CPV effective action
may arise from elsewhere, from summed small half-circle integrals around pole terms.
The contour integral parallel to the positive real axis
may be deformed to the line along the imaginary axis,
$0 \sim - i \infty$.
This deformation picks up infinitely many contour integrations
around poles at $t = e^{-i \pi/4} n \pi/ (c, d)\,, n = 1,2, \cdots$.
This can be seen more clearly by an equivalent integrand formula,
\begin{eqnarray}
&&
%\hspace*{1cm}
{\cal M}^{\rm CPV} ({\cal F}, {\cal G}; t) =  -   t^2 {\cal G} +
\nonumber \\ &&
\frac{i t^2}{2}  {\cal F} \left( \cot ( t c \, e^{i\pi/4} ) \coth  ( t d\,  e^{i\pi/4} )
\right.
\nonumber \\ &&
\left.
- 
\cot (t d\,  e^{i\pi/4} ) \coth  (t c\, e^{i\pi/4} )
\right) 
\,.
%\nonumber \\ &&
\end{eqnarray}
Summing up all pole terms, one arrives at the CPV effective action,
\begin{eqnarray}
&&
\Delta {\cal L}_p^{\rm CPV} = - \frac{{\cal F}}{8 \pi^2}
\sum_{n=1}^{\infty} \frac{(-1)^n}{n} \times
\nonumber \\ &&
\left(\exp[- \frac{m^2 n\pi}{ \sqrt{2} |c| }(1 - i) ] \coth \frac{ d n \pi}{c }
\right.
\nonumber \\ &&
%\hspace*{0.5cm}
\left.
-
\exp[- \frac{m^2 n\pi}{ \sqrt{2} |d| }(1 - i) ] \coth \frac{ c n \pi}{d }
\right)
\,.
\end{eqnarray}

For definiteness we shall assume a positive ${\cal G}$, which implies
$d ({\cal F}, {\cal G})> c ({\cal F}, {\cal G})$.
One can use in the above formula $\coth x = 1 + 2 e^{-2x}/(1- e^{-2x})$
with $x= n \pi d/c$ or $x= n \pi c/d$.
Let us assume $d/c \gg 1$ which leads to
\begin{eqnarray}
&&
\Delta {\cal L}_p^{\rm CPV} \approx - \frac{{\cal F}}{8 \pi^2}
\sum_{n=1}^{\infty} \frac{(-1)^n}{n} \times
\nonumber \\ &&
\hspace*{-0.5cm}
\left(\exp[- \frac{m^2 n\pi}{ \sqrt{2} |c| }(1 - i) ]
-
\frac{2}{ n\pi} \frac{c }{ d} \exp[- \frac{m^2 n\pi}{ \sqrt{2} |d| }(1 - i) ] 
\right)
\,.
\end{eqnarray}
The power series here can be analytically calculated, to give
\begin{eqnarray}
&&
\Delta {\cal L}_p^{\rm CPV} \approx  - \frac{1}{24 \pi} \frac{{\cal F}^2}{{\cal G} }
%\nonumber \\ && \hspace*{-0.5cm}
+
\frac{{\cal F}}{8 \pi^2} 
\ln \frac{m^2 \pi (1-i) + \frac{{\cal F}} {\sqrt{2 {\cal G}} } }{m^2 \pi (1-i) }
\,.
%\nonumber \\ &&
\end{eqnarray}
Strictly, this contains ${\cal G}$ even terms, which should be dropped.
In the small mass limit the second term dominates:
\begin{eqnarray}
&&
\Delta {\cal L}_p^{\rm CPV} \approx  \frac{{\cal F}}{16 \pi^2} 
\ln \frac{{\cal F}^2 }{ m^4 {\cal G}}
\,.
\end{eqnarray}
In the large mass limit the first term $\sim - \frac{{\cal F}^2}{24 \pi {\cal G}} $ dominates.

The presence of inverse ${\cal G}-$powers in these results is annoying,
and it is better to use the direct integral formula along the real $t$ axis
of (\ref{cpv m-function 2}), in which we encounter no singularity.

\subsection{Light neutrino one-loop contribution}

From results of the preceding subsections,
the neutrino one-loop effective action occurs under Z-field background.
Its relevant squared gauge coupling is $(g^2 + (g')^2)/4 $.

As to the nature of neutrino mass types there are two possibilities,
of Dirac type and of Majorana type.
We assume that light Majorana neutrinos are generated
by the seesaw mechanism \cite{minkowski} due to
the mixing of Dirac type Higgs coupling with heavy right-handed neutral
Majorana leptons, necessarily SU(2) $\times $ U(1) gauge singlet.
Following Appendix D, one can deal with the Majorana type without
much difficulty.
Calculations are the same for both types of neutrino masses, and
the difference appears in the lepton number.
As is well known, the Majorana neutrino has an indefinite lepton number,
or rather one can assign its lepton number $+1$.
The energy shift and the decay rate related to the real and the imaginary parts
of effective action depend on the absolute value of neutrino masses and not
on its mass type.

We focus on a field value region larger than $m_{\nu}^2/e$ with $e= \sqrt{4\pi \alpha}$. 
Numerically, this is the region of field strength,
\begin{eqnarray}
&&
\sqrt{| {\cal F}|} \,, \sqrt{| {\cal G}|} > \frac{ m_{\nu}^2}{e}
 \sim 170 \, {\rm mV cm}^{-1} (\frac{m_{\nu}}{{\rm meV}})^2
\,.
\end{eqnarray}
The relevant mathematical limit is then $m_{\nu} \rightarrow 0$,
and one derives, following the calculation in Appendix A
that led to (\ref{massless limit math}),
\begin{eqnarray}
&&
\Im {\cal M}_{\nu} ({\cal F}_Z, {\cal G}_Z; s) \approx \frac{ g^2 + (g')^2}{4}
\frac{{\cal G}_Z }{8\pi^2} \ln \frac{|a_Z| t_0 }{ m_{\nu}^2 \pi }
\,,
\\ &&
a_Z^2 =  {\cal F}_Z + \sqrt{{\cal F}_Z^2 + {\cal G}_Z^2 }
\,,
\end{eqnarray}
with $t_0$ of order unity dimensionless number.
Three neutrino species (mass eigenstates) add with the same sign.

The real part of massive fermion loop is given by 
\begin{eqnarray}
&&
\Re \Delta {\cal L}_f = \frac{ a b}{4\pi^3} \sum_{n=1}^{\infty} \frac{1}{n}
\nonumber \\ &&
\left( \coth \frac{a n\pi }{ b} I_1 (\frac{m^2 n \pi x}{|b|}  )
- \coth \frac{b n\pi }{ a} I_2 (\frac{m^2 n \pi x}{|a|} )
\right)
\,,
\\ &&
I_1(\mu_n) = \int_0^{\infty} dx\, \frac{x}{x^2 + 1} e^{-\mu_n x}
\,, \hspace{0.3cm}
\mu_n = \frac{m^2 n \pi x}{|b|}
\,,
\\ &&
\hspace*{-0.3cm}
I_2(\mu_n') = \int_0^{\infty} dx\, \frac{x}{x^2 + 1} \cos (\mu_n' x)
\,, \hspace{0.3cm}
\mu_n' = \frac{m^2 n \pi x}{|a|}
\,.
\end{eqnarray}
The dominant term in the massless limit is
\begin{eqnarray}
&&
\hspace*{-0.5cm}
\Re \Delta {\cal L}_f \approx \frac{ a b}{4\pi^3} \sum_{n=1}^{\infty} \frac{1}{n}
%\nonumber \\ &&
\left(  I_1 (\frac{m^2 n \pi x}{|b|} ) -I_2 (\frac{m^2 n \pi x}{|a|} )
\right)
\,.
\end{eqnarray}
Hence, discussion of the massless limit $\mu \rightarrow 0$ depends on
the limiting behaver of two integrals, $I_i (\mu)$.

\begin{figure*}[htbp]
 \begin{center}
 \epsfxsize=0.4\textwidth
 \centerline{\epsfbox{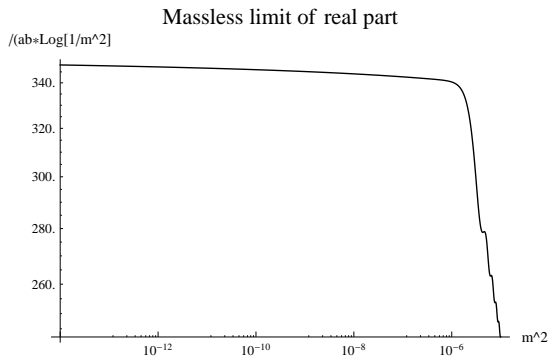}} \hspace*{\fill}\vspace*{1cm}
   \caption{
Behavior of the real part of neutrino one-loop contribution
divided by $\ln (1/m_{\nu}^2)$ factor.
}
   \label {small mass nu limit}
 \end{center} 
\end{figure*}

We attempted to derive the massless limit of the real part of effective lagrangian
using analytic method similar to the one in Appendix A,
but failed, presumably under a bad control of interchangeable operations.
Numerical simulations, as illustrated in Fig(\ref{small mass nu limit}),
however, show that the limiting behavior follows the logarithmic 
singularity $\sim \ln (1/m^2)$ as $m \rightarrow 0$.

\section{Electroweak theory under field background at finite temperature} 

\subsection{Effective action under background gauge fields at finite temperature}

The usual procedure from the real time quantum field theory to quantum field theory
in thermal equilibrium is to use the Euclidean field theory in a finite Euclidean time,
$0 \sim - i \beta = - i /T$ \cite{finite temp field theory} 
with $T$ the temperature in the unit of Boltzmann constant taken to be unity.
It is most unambiguous to go back to the starting equation of Schwinger
in the real-time formalism,
\begin{eqnarray}
&&
{\cal L}^{(1)}(x) = \frac{i}{2} \int_0^{\infty} \frac{ds}{s} e^{- i m^2 s} {\rm tr}
\langle x | \exp[- i ( \Pi^2 - \frac{1}{2} \sigma \cdot F ] |x \rangle
\,.
\nonumber \\ &&
\end{eqnarray}
The proper-time $s$ here has $-2$ mass dimension, while it is $-1$
in the usual approach.
It is legitimate to replace $ms $ by $- is \beta = - is/T $
for finite-temperature quantum field theory \cite{finite temp field theory}.
Hence the finite Euclidean interval should be $0 \sim - i /(m T) $
for transformation from the Schwinger proper-time approach.
Our ansatz of background field theory under constant gauge fields is thus
to use the integration path,
\begin{eqnarray}
&&
\hspace*{-0.5cm}
\frac{i}{2} \int_0^{-i \beta/m}  \frac{ds}{s} e^{- i m^2 s} {\rm tr}
\langle x | \exp[- i ( \Pi^2 - \frac{1}{2} \sigma \cdot F) ] |x \rangle
\,,
\end{eqnarray}
for $ {\cal L}^{(1)}(\vec{x}, T)$.

It is useful to rescale the Euclidean proper-time by
introducing dimensionless Euclidean time and making Wick rotation.
We shall present the case of fermion one-loop.
Results for the effective action per unit space-volume $\Delta {\cal W}_f^{(T)}/V$
are for the vector contribution,
\begin{eqnarray}
&&
\frac{( m T)^{3/2} }{16\pi^2} \int_0^1 \frac{d\tau }{\tau^3}e^{i m \tau/T} {\cal M}_V
({\cal F} ,  {\cal G};  \frac{\tau}{mT})
\,,
\end{eqnarray}
and for the axial-vector contribution,
\begin{eqnarray}
&&
\frac{ ( m T)^{3/2} }{16\pi^2} \int_0^1 \frac{d\tau }{\tau^3}e^{i m \tau/T} {\cal M}_A
({\cal F} ,  {\cal G};  \frac{\tau}{mT} )
\,.
\end{eqnarray}
Relation between the axial-vector and the vector cases is the same as
before: $- {\cal F} \leftrightarrow i {\cal G}$ rule holds.
For example,
the same equations for ${\cal M}$ such as (\ref {cpv m-function 1}) $\sim$
(\ref {cpv m-function 4}) may be used provided the integration
variable is replaced by $t \rightarrow - i \frac{\tau}{m T} $.

Extension to scalar and gauge boson loops at finite temperature should be evident.

\subsection{Massless limit at finite temperature}

The massless limit of fermion one-loop contribution to the effective lagrangian
is mathematically equivalent to
zero temperature limit, since various ${\cal M}$ 
functions at finite temperature depends on 
 of dimensionless variable in the combinations,  ${\cal F}/(m T)^2\,, {\cal G}/(mT)^2$.
General theory of finite temperature perturbation theory
\cite{finite temp field theory} suggests that
the zero temperature limit should coincide with
the usual effective action without temperature,
which is shown above to exhibit the logarithmic singularity $\sim \ln (1/m^2)$.

Whether this argument is correct or not
is not evidently clear  under the presence of background gauge fields, however.
Let us look into one of typical finite temperature formula in our setting:
\begin{eqnarray}
&&
\hspace*{-0.3cm}
I = \frac{(mT)^{3/2}}{16 \pi^2} \int_0^1 \frac{d\tau}{\tau^3}\, e^{im \tau/T}
{\cal M} (\frac{c({\cal F}, {\cal G})}{m^2 T^2} \tau^2\,, \frac{d({\cal F}, {\cal G})}
{m^2 T^2}\tau^2 )
\,,
%\nonumber \\ &&
\\ &&
{\cal M}(x, y) = i xy \frac{ (1 + i \tanh \frac{x}{\sqrt{2}}  \tan \frac{x}{\sqrt{2}} ) 
(1 - i \tanh \frac{y}{\sqrt{2}}  \tan \frac{y}{\sqrt{2}} )}
{ (\tanh \frac{x}{\sqrt{2}} + i \tan \frac{x}{\sqrt{2}} ) 
( i \tanh \frac{y}{\sqrt{2}} +  \tan \frac{y}{\sqrt{2}} ) } 
\nonumber \\ &&
- 1 - \frac{i}{3} (x^2 - y^2)
\,,
\end{eqnarray}
where $ c({\cal F}, {\cal G}), d({\cal F}, {\cal G})$ are given by (\ref{cpv m-function 3})
and (\ref{cpv m-function 4}).
One can separate $\tau-$integration range into two parts,
one near $\tau=0$ and the rest.
The former contribution is relevant only for large values of the mass,
and is irrelevant to discussion here.
In the latter contribution  the large $(x,y)$ variable limit
of $ {\cal M}(x, y)$ can be taken, to use an approximate formula,
\begin{eqnarray}
&&
{\cal M}(x, y) \approx xy - \frac{i}{3} (x^2 - y^2)
\,.
\end{eqnarray}
Integration near the upper boundary $\tau=1$ gives
\begin{eqnarray}
&&
I \rightarrow \frac{(mT)^{-5/2}}{16 \pi^2} \int_0^1 d\tau\, \tau \,  e^{im \tau/T}
\times
\nonumber \\ &&
\left( c({\cal F},{\cal G}) d({\cal F},{\cal G}) 
 - \frac{i}{3} (c^2({\cal F},{\cal G})  - d^2({\cal F},{\cal G}) \,)
\right)
\,.
\label {m0 limit}
\end{eqnarray}
Thus, the limiting behavior as $m \rightarrow 0$ is
\begin{eqnarray}
&&
\hspace*{-0.5cm}
I \rightarrow O( (mT)^{-5/2} ) %\times
%\nonumber \\ && 
\frac{1}{32 \pi^2} \left( {\cal F}^2
+ \frac{2 i}{3} {\cal G} \sqrt{ {\cal F}^2 + {\cal G}^2}
\right)
\,.
%\nonumber \\ &&
\end{eqnarray}
The second term of this equation is CP violating,
since 
\begin{eqnarray}
&&
c^2 - d^2 = 4 {\cal G}_Z \sqrt{{\cal F}_Z^2 +{\cal G}_Z^2 }
\,,
\end{eqnarray}
is ${\cal G}_Z$-odd.
Relevant gauge coupling is $g^2 + (g')^2 = e^2 /(\sin^2 \theta_w \cos^2 \theta_w)$, and
the entire limit function $I_0$ has the form, recovering
gauge coupling constants,
\begin{eqnarray}
&&
I_0 = \frac{(g^2 + (g')^2 )^2 }{ 32 \pi^2} \left(
{\cal F}_Z^2 - i \frac{4}{3} {\cal G}_Z \sqrt{{\cal F}_Z^2 +{\cal G}_Z^2 }
\right)
\nonumber \\ &&
\times
\sum_{\nu} ( m_{\nu} T)^{-5/2} f(\frac{T}{m_{\nu}})
\,,
\label {snu-mass limit}
\\ &&
f(x) = \int_0^1 d\tau\, \tau \,  e^{i x \tau} = \frac{-1 + e^{ix} (1 - i x)}{x^2}
\,.
\end{eqnarray}
The function $f(x)$ here has both the real and the imaginary parts.

Both the real and the imaginary parts of effective lagrangian $I_0$ contain
 sinusoidal behaviors in the 
temperature-related variable $m_f/T$.
This exhibits slowly varying oscillatory components with small neutrino masses.
Their oscillation time period $t_p$ at time $t$ is estimated by
\begin{eqnarray}
&&
\frac{t_p}{t} = 4 \pi \frac{T}{m_{\nu}}
\,.
\end{eqnarray}
We  speculate that
a large time-period implied by a small neutrino mass may have observational impact 
at the relevant epoch of cosmological evaluation of $t_p/t = O(1)$.

It is  useful here to explain how CP violating observables may
emerge from CP conserving initial states.
Take as an example the formula given by (\ref{snu-mass limit}) and
focus on the real part of the second contribution.
The combination $ {\cal G}_Z \sqrt{{\cal F}_Z^2 +{\cal G}_Z^2 }$
involves four external gauge fields, and one may take
two of them as initial state and the two other as final state.
When one takes $\sqrt{{\cal F}_Z^2 +{\cal G}_Z^2 }$ for the initial state,
it is CP conserving, while the final one ${\cal G}$ is CP violating.
This means that the final CPV state is created from an initial CPC
state.
The real part of this effective lagrangian gives the transition amplitude
for this CPV process.
What is surprising here is that this CPV transition rate is enhanced by
a large factor $\propto (m_{\nu} T)^{-5}$ due to the small neutrino mass.

It is conceivable to think of mixed background made of
electromagnetic fields and Z-fields, considering electron one-loop.
In this case CP-even electromagnetic field can create
CPV transition.
For example, high intensity laser collision can create
CPV effect, although rate of this mixed case is suppressed
by inverse powers of electron mass.

\subsection{High temperature limit and a mechanism of generating lepton number asymmetry}

High temperature expansion gives the leading contribution to CP conserving (CPC)
and CP violating (CPV) effective action of the form,
\begin{eqnarray}
&&
\frac{\Delta W^{CPV}}{V} \approx \frac{g_A^6}{2520 \pi^2} \frac{{\cal G}}{m^8 \sqrt{m T} }
( 13 {\cal F}^2 + 8 {\cal G}^2 ) \left( 1 - e^{i m/T} \right)
\,,
\nonumber \\ &&
\\ &&
\frac{\Delta W^{CPC}}{V} \approx - \frac{1}{1440 \pi^2} \frac{{\cal S}}
{m^4 \sqrt{m T}  } \left( 1 - e^{i m/T} \right)
\,,
\\ &&
{\cal S} = (7 g_A^4- 4 g_V^4) {\cal F}^2 + (4 g_A^4 - 7 g_V^4 ){\cal G}^2
\,,
\end{eqnarray}
recovering gauge coupling constants of vector and axial-vector parts, $g_V\,, g_A$.
High temperature expansion is equivalent to small field expansion, since
these dependence appears via the combination, ${\cal F}/(m T)^2\,, {\cal G}/(m T)^2$
in various ${\cal M}$ functions.

Flavor dependence of the  effective lagrangian is as follows:
\begin{eqnarray}
&&
\hspace*{1cm}
\Delta {\cal L}^{\rm CPV} = \frac{ 1}{ 2520 \pi^2} \times
\nonumber \\ &&
\sum_f
\left( (\frac{g}{2})^6 \frac{ {\cal G}_W}{\sqrt{m_f T} } (13 {\cal F}_W^2 + 8 {\cal G}_W) 
 \frac{1}{m_f^8 } (1 - e^{i m_f/T} )
\right)
\nonumber \\ &&
%\hspace*{-0.5cm}
+ \sum_f 4 (g^2+(g')^2\,)^3 
\frac{ {\cal G}_Z}{\sqrt{m_f T} } (13 {\cal F}_Z^2 + 8 {\cal G}_Z)
\nonumber \\ &&
 \times
 \frac{f_L^6 - f_R^6}{m_f^8 } (1 - e^{i m_f/T} )
\,,
%\nonumber \\ &&
\\ &&
\hspace*{1cm}
\Delta {\cal L}^{\rm CPC} = - \frac{ 1}{ 1440 \pi^2} \times
\nonumber \\ &&
\sum_f 
\left[
\left( 3 (\frac{g}{2})^4 \frac{ {\cal F}_W^2 - {\cal G}_W^2 }{ \sqrt{m_f T}} 
\frac{ 1 - e^{i m_f/T}}{m_f^4 }
\right)
\right.
\nonumber \\ &&
\hspace*{-0.5cm}
\left.
+ 
\sum_f
\left( (g^2 + (g')^2\, ) \frac{ {\cal F}_Z^2 - {\cal G}_Z^2 }{ \sqrt{m_f T}} 
\right)
 \frac{f_L^4 - f_R^4 }{m_f^4 } (1 - e^{i m_f/T} )
\right]
\,.
\nonumber \\ &&
\end{eqnarray}
We neglected electromagnetic field background contributions.
$f_L, f_R$ are values listed in (\ref{z f-factor 1}) $\sim $ (\ref{z f-factor 4}).

If the neutrino mass is of Majorana type, the lepton number is violated
by Z-field fluctuation.
Along with CP violation discussed here, the combined symmetry violation
may lead to a mechanism of generating lepton number asymmetry.
A missing out-of equilibrium condition \cite{sakharov} can be replaced
by a generation of the lepton number chemical potential \cite{cohen-kaplan}
due to the lepton number violating scattering involving right-handed 
heavy Majorana $N_R$ \cite{my 2022}.
The lepton asymmetry is not converted to a baryon number
since the generation occurs at epochs much later than
sphaleron formation \cite{sphaleron}
unlike the usual lepto-genesis scenario \cite{fukugita-yanagida}. 
The lepton asymmetry thus generated may however lead to
a large neutrino degeneracy left to the present cosmic epoch.

\subsection{Speculation on a possible phase transition}

Skeptics might have detected a peculiar behavior  of approach to zero temperature.
The behavior $\propto (mT)^{-5/2}$ is inconsistent with a smooth
approach to zero temperature having a milder logarithmic singularity on the neutrino mass.
Namely, a discontinuity at zero temperature is present.
The only way we can think of for a proper behavior  is to 
assume a phase transition at some critical temperature
$T = T_c$ or $\beta = \beta_c$, and our result so far is applicable
only in the higher temperature region  $T > T_c$.

The effective potential (= $-$ the effective lagrangian) is
identified to $-$ the free energy in many-body system in thermal equilibrium.
We postulate that the free energy behaves as
\begin{eqnarray}
&&
m_{\nu}^{-5/2} (T- T_c)^{-5/2} \alpha^2 ({\cal F}, {\cal G})^2 \theta (T - T_c)
\,,
\end{eqnarray}
with $\alpha$ the fine structure constant
near and above the critical temperature. We neglected all complications of
order unity. We leave the appropriate combination of field strength unspecified.
There are two possibilities on how to equate the discontinuous quantity,
\begin{eqnarray}
&&
(m_{\nu}T_c)^{-5/2} \alpha^2 ({\cal F}, {\cal G})^2
\,,
\end{eqnarray}
to:
the first is to take $T_c^3$ and the second is to choose $(m_{\nu} T_c)^{3/2}$.
We believe the first possibility more likely, and present the result in this case.
The estimated critical temperature is
\begin{eqnarray}
&&
T_c = m_{\nu}^{-5/11} ( \alpha |({\cal F}, {\cal G})| )^{4/11}
\,.
\end{eqnarray}
For field strength $\geq m_{\nu}^2$,
the critical temperature $T_c \geq \alpha^{4/11} m_{\nu}$.

In the case of Dirac neutrino one could add another order parameter,
the chemical potential $\mu_{\nu}$, to make the phase diagram
in the three parameter space $(T, \mu_{\nu})$ and field strength.
We need a separate study to analyze the phase space diagram
in a more rigorous way.

\section{Summary and outlook}

We extended the gauge invariant proper-time background field method
to chiral gauge theories which makes possible to derive the
effective action for the standard electroweak theory.
Result in a chiral gauge theory is obtained by using a simple
transformation rule from the corresponding vector-like gauge theory.
The rule for the integrand function ${\cal M}({\cal F}, {\cal G})$
is given by (\ref{axial from vector}) where the right-hand side
is from the corresponding vector-like theory, (\ref{qed in a,b}), (\ref{qed mf}).
We applied this method both at zero and finite temperature.

We found interesting results on the infrared singularity in the massless 
neutrino limit, and identified a new source of CP violation
in addition to the well-known and established Kobayashi-Maskawa
CP violating phase of quark mass mixing matrix.
CPV contribution at finite temperature is very strong behaving 
$\propto (m_{\nu} T)^{-5/2}$  in the small neutrino mass limit.
The  limit  formula of the effective action is
given by (\ref{snu-mass limit}).

In the last section we speculated on the phase transition
that suggests existence of a shifted critical temperature
under gauge field background.
A more elaborate analysis is required to determine whether
this possibility is real.

Search for physical consequences within a observational reach
is yet to come.
We can list an important unsolved problem ahead of us:
lack of a reliable estimation to calculate gauge field backgrounds
that may exist in problems at our hand.
The problem may be solved in cosmology at high temperatures
by some elaborate technique.
Cases at zero temperature do depend on special features
of the problem, and we need to work out individual cases.
The possibility of coherence generation is necessary to discuss whether
condensate formation of gauge fields becomes possible.
The condensation may be realized deep interior of neutron stars.

Despite of missing concrete application we hope that this work laid down some
fundamentals to a new area of research activity.

\section{Appendix: Additional notes}

\subsection{Logarithmic singularity in the massless fermion limit}

We prove that the imaginary part given by eq.(\ref{imaginary part of f-loop})
leads to the logarithmic singularity of the type $\propto \ln (1/m^2$.

The first step is to use the identity,
\begin{eqnarray}
&&
\coth A_n = 1 + 2 \frac{e^{-2A_n}}{1 - e^{-2A_n}}
\,, \hspace{0.5cm}
A_n = \frac{b n \pi}{a}
\,,
\end{eqnarray}
and observe the power series in (\ref{imaginary part of f-loop})
from the second term to converge after the interchange of the limit $m \rightarrow 0$
and the sum.
This means that one can replace $\coth A_n$ by the unity $1$ for the discussion
of the infrared limit,
\begin{eqnarray}
&&
\lim_{m \rightarrow 0} \Im \Delta {\cal L}_f = \frac{ab}{8\pi^2}
\lim_{m \rightarrow 0} 
\sum_{n = 1}^{\infty} \frac{e^{- m^2 n \pi/|a| }  }{n}
\,.
\end{eqnarray}
The problem thus reduces to the limit of a special case $\phi(1, e^{- m^2 n \pi/|a|})$
called Appell function,
\begin{eqnarray}
&&
\phi (z, s) = \sum_{n=1}^{\infty} \frac{s^n}{n^z}
\,.
\end{eqnarray}
According to \cite{whittaker-watson},
this function has an integral representation,
\begin{eqnarray}
&&
\phi(1, s) = s \int_0^{\infty} dt\, \frac{1}{e^t - s}
\,, \hspace{0.5cm}
s = \exp[- \frac{m^2 \pi }{|a|}]
\,.
\end{eqnarray}

In the massless limit one can approximate $s$ by
\begin{eqnarray}
&&
s \approx 1 - \epsilon
\,, \hspace{0.5cm}
\epsilon = \frac{m^2 \pi }{|a|}
\,,
\end{eqnarray}
to give
\begin{eqnarray}
&&
\phi(1, s) \approx  \int_0^{\infty} dt\, \frac{1}{e^t - 1 + \epsilon}
\,.
\end{eqnarray}
The integrand has a pole outside the integration region at
$t = \ln ( 1- \epsilon) \sim - \epsilon < 0$.

A standard estimation of this integral in the small $\epsilon$ limit would be
to separate the integral into two parts, the one $[0, t_0]$ and the other $[t_0, \infty]$
taking $t_0$ of order unity $\gg \epsilon$.
The second contribution is $\epsilon-$independent and may be dropped
in the limit.
The integration of the first contribution near the end point $0$ is crucial,
and one may approximate this integral to derive
\begin{eqnarray}
&&
\hspace*{-0.5cm}
\lim_{m \rightarrow 0}  \phi(1, s) \approx \int_0^{t_0} dt\, \frac{1}{t + \epsilon}
= \ln \frac{t_0 + \epsilon}{\epsilon} \rightarrow \ln \frac{ |a| t_0}{ m^2 \pi}
\,.
\label {massless limit math}
\end{eqnarray}
One cannot determine $t_0$ precisely, but this is irrelevant
to the logarithmic behavior.

We checked this derivation numerically as well.

\subsection{Color matrix eigenvalues}

The problem is to derive $z$ eigenvalues of the algebraic equation
for SU(3) color,
\begin{eqnarray}
&&
{\rm det} \left( z \delta_{ab} - 
\sum_{i =1}^8 \frac{(\lambda_i)_{ab}}{2} t_i \right) = 0
\,,
\end{eqnarray}
where $\lambda_i$ is $3\times 3$ hermitian Gell-Mann matrices,
and $t^i$'s are real numbers.

To simplify the equation, we first use the unitary transformation such that
the matrix part is diagonal, taking  only $t_3\,, t_8$ non-vanishing.
The third order algebraic eigenvalue equation is then
\begin{eqnarray}
&&
f_3(z) = z^3 - \frac{1}{4}(t_3^2 + t_8^2) + \frac{1}{12 \sqrt{3}} t_8 ( t_8^2 - 3 t_3^2)
\nonumber \\ &&
= \left( \frac{ t_3^2 + t_8^2}{4} \right)^{3/2} ( x^3 - x + c)
\,, 
\\ &&
x = \frac{2 z }{(t_3^2 + t_8^2 )^{1/2} }
\,, \hspace{0.5cm}
c = \frac{1}{6 \sqrt{3}} \frac{t_8  ( t_8^2 - 3 t_3^2)}{ (t_3^2 + t_8^2 )^{3/2}}
\,.
\end{eqnarray}
$c$ is an odd function of $t_8$, and ranges between $- 0.0 962 \sim 0.0962$
with extrema given at $t_8/t_3 = \pm 0.57735$.
At $t_8/t_3 =\infty$ it  goes to the limiting value $1/(6\sqrt{3})=0.0962 $.

The eigenvalue equation thus reduces to solving
\begin{eqnarray}
&&
g_3(x) = 0
\,, \hspace{0.5cm}
g_3(x) =
x^3 - x + c
\,,
\end{eqnarray}
where $c$ is a function of  single variable combination, $t_8/\sqrt{t_3^2 + t_8^2} $. 
The function $x^3 - x$ has two local extrema at $x= \pm 1/\sqrt{3} \sim \pm 0.5774$:
local maximum $2/(3\sqrt{3}) \sim 0.3849 $ at $x= - 1/\sqrt{3}$ and local minimum 
$- 2/(3\sqrt{3}) \sim - 0.3849 $ at $x=  1/\sqrt{3}$.
Hence it is easy to see that three real zeros of $x$ exist:
two positive and one negative for $t_8 < 0$ and one positive and two negative for $t_8>0$.
At $t_8 = 0$ three eigenvalues are $x = 0\,, \pm 1$, hence
$z = 0\,, \pm \frac{t_3}{2}$. 
In general eigenvalues are functions of $\sqrt{t_3^2 + t_8^2}/2 \times (-1 \sim 1)$.
Prior to the diagonalization they are the Casimir invariant $\sum_i \sqrt{t_i^2}/2 \times$
numbers of absolute value less than or equal to unity.

The eigenvalue problem of SU(2) group is simpler and its solution is trivially derived.

\subsection{Scalar and gauge boson loops}

For completeness we record known relevant functions, ${\cal M}_s$ for scalar
 and ${\cal M}_g$ for gauge boson \cite{yildiz-cox}, \cite{claudson-yildiz-cox},
that replace the fermion ${\cal M}_f$ 
in the same integral formula of $\Delta {\cal L}$, eq.(\ref{qed in a,b})
with appropriate choice of mass in $e^{-m^2 s}$:
\begin{eqnarray}
&&
\hspace*{-0.5cm}
{\cal M}_s = - \frac{1}{4} \left( \frac{s^2 a b}{ \sin (sa) \sinh (sb)} - 1 - \frac{s^2}{6}
(a^2 - b^2) \right)
\,,
\\ &&
{\cal M}_g = \frac{s^2 a b}{ \sinh (sa) \sin (sb)} (\sinh^2(sa) - \sin^2(sb) \,) 
\nonumber \\ &&
- s^2 (a^2 - b^2)
\,.
\end{eqnarray}
There is no massless singularity in the effective action calculated
from ${\cal M}_s$ and ${\cal M}_g$.

\subsection{General four-component neutral lepton system}

It is best to introduce four-component spinor field
incorporating two 2-component $\nu_L\,, N_R$, and discuss
the seesaw mechanism in general by introducing three independent
mass mixing terms.
Some elementary explanation of Majorana fermion is given in the textbook of
Aitchinson-Hey \cite{ah textbook}.
Quantization based on two-component field is explained in \cite{my 07}.

One introduces creation and annihilation operators, both for two-component
fields, denoted by $c_p^{\dagger}\,, c_p \,, d_p^{\dagger}\,, d_p$.
Quantum field projected on plane waves is described by
\begin{eqnarray}
&&
\psi (x; p) = e^{-i p \cdot x} 
\left(
\begin{array}{c}
u(p) c_p  \\
i \sigma_2 v(p)^* d_p^{\dagger}   
\end{array}
\right) + e^{i p \cdot x} 
\left(
\begin{array}{c}
v(p) d_p \\
i \sigma_2 u(p)^* c_p^{\dagger} 
\end{array}
\right)
\,,
\nonumber \\ &&
\end{eqnarray}
with $u\,, v$ two-spinors.
General $4 \times 4 $ mass matrix is introduced;
\begin{eqnarray}
&&
{\cal M} = 
\left(
\begin{array}{cc}
m_L & m_D   \\
 m_D & m_R
\end{array}
\right)
\,.
\end{eqnarray}
For simplicity we assume real $m_i$.
In the usual SU(2) $\times $ U(1) theory $m_L = 0$.
Note that our $\gamma$ matrices are not conventional
in order to make $\gamma_5$ diagonal, in particular, $\gamma^0$
is off-diagonal, and
\begin{eqnarray}
&&
\gamma^i =
\left(
\begin{array}{cc}
0 &  \sigma_i  \\
- \sigma_i  & 0
\end{array}
\right)
\,, \hspace{0.3cm}
\gamma^0 =
\left(
\begin{array}{cc}
0 &  1  \\
1 & 0
\end{array}
\right)
\,, 
\\ &&
\gamma^5 = i \gamma^0 \gamma^1 \gamma^2 \gamma^3 =
\left(
\begin{array}{cc}
-1 &  0  \\
0 & 1
\end{array}
\right)
\,.
\end{eqnarray}

The mass and the kinetic terms are defined  as usual, giving
\begin{eqnarray}
&&
%\hspace*{-1cm}
\bar{\psi} {\cal M} \psi = m_L \, c_p^{\dagger} c_p  
\left( u(p)^{\dagger} ( i \sigma_2 ) u(p)^* -  u(p)^T  ( i \sigma_2 ) u(p)
\right) 
\nonumber \\ &&
+  m_R\, d_p^{\dagger} d_p
\left( v(p)^{\dagger} ( i \sigma_2 ) v(p)^* -  v(p)^T  ( i \sigma_2 ) v(p)
\right) 
\nonumber \\ &&
+ m_D \, \left (c_p^{\dagger} d_p \, u(p)^{\dagger} v(p) + 
d_p c_p^{\dagger} \, v(p)^T u(p) \right)
\,,
\\ &&
\bar{\psi} (- i \partial \cdot \gamma) \psi =
2 c_p^{\dagger} c_p \, u(p)^{\dagger} (p_0 -\vec{p}\cdot \vec{\sigma} ) u(p)
\nonumber \\ && 
+ 2\,
d_p^{\dagger} d_p \, v(p)^{\dagger} (p_0 -\vec{p}\cdot \vec{\sigma} ) v(p)
\,.
\end{eqnarray}
We neglected highly oscillating terms $\propto e^{\pm 2i p \cdot x}$
and irrelevant c-number terms.
The first two mass terms $\propto m_L\,, m_R$  give Majorana type masses, and
the last $\propto m_D$  is of Dirac type.
In the kinetic term $\gamma^0 \gamma \cdot \partial = \partial_0 + \gamma_5 \vec{\nabla}\cdot \vec{\Sigma} $, a $2\times 2$ block-diagonal form.
Two kinetic terms correspond to chirally projected kinetic terms,
$(1 \pm \gamma_5)/2$.

In the chiral basis of $(1 \mp \gamma_5)/2$ Majorana mass terms
are diagonal, while Dirac terms are off-diagonal, causing a mixing 
between two different chirality states.
This gives rise to the seesaw mechanism \cite{minkowski}.
Suppose a hierarchical parameter pattern, $m_R \gg m_D \gg m_L$
in the limit $m_L \rightarrow 0$.
Diagonalization of the mass matrix ${\cal M}$ leads to
eigenvalues and their eigenvectors of the form,
\begin{eqnarray}
&&
m \approx - \frac{m_D^2}{m_R}\,, \; {\rm state\;}
| m \rangle \sim | \nu_L \rangle \mp \frac{m_D}{m_R} | N_R \rangle
\,, 
\\ &&
M  \approx m_R
\,, \; {\rm state\;}
| M \rangle \sim | N_R \rangle \pm \frac{m_D}{m_R} | \nu_L  \rangle
\,.
\end{eqnarray}
By a phase change one can take $m$ positive.

Thus,  the four-component chiral  formalism can
be constructed in terms of the chiral basis
 by assuming a small Majorana mass $m$ for
chiral $\sim |\nu_L \rangle$ state and a large Majorana mass $M$
for $\sim |N_R \rangle $ state.
The chiral projection of $(1 -\gamma_5)/2$ makes calculation in the text for Dirac neutrinos
equally applicable to light Majorana neutrinos generated by
the seesaw mechanism,
giving the same energy shift and the same decay rate.

There is, however, an important difference with regard to
lepton number violation:
the real neutrino-pair production process of its rate 
given by the imaginary part of effective lagrangian 
is lepton-number violating, and CP-violating.

\vspace{0.5cm}
\begin{acknowledgments}
% put your acknowledgments here.
This research was partially
 supported by Grant-in-Aid   21K03575   from the Japanese
 Ministry of Education, Culture, Sports, Science, and Technology.

\end{acknowledgments}

\end{document}